\documentclass[sigconf,natbib=true]{acmart}
\usepackage{multirow}
\usepackage{colortbl}

\usepackage{bm}

\def\eqref#1{equation~\ref{#1}}

\def\1{\bm{1}}

\def\va{{\bm{a}}}
\def\vb{{\bm{b}}}

\def\ve{{\bm{e}}}

\def\vg{{\bm{g}}}
\def\vh{{\bm{h}}}

\def\vp{{\bm{p}}}
\def\vq{{\bm{q}}}

\def\vs{{\bm{s}}}

\def\vz{{\bm{z}}}

\def\mH{{\bm{H}}}

\def\mK{{\bm{K}}}

\def\mQ{{\bm{Q}}}

\def\mV{{\bm{V}}}
\def\mW{{\bm{W}}}

\DeclareMathAlphabet{\mathsfit}{\encodingdefault}{\sfdefault}{m}{sl}
\SetMathAlphabet{\mathsfit}{bold}{\encodingdefault}{\sfdefault}{bx}{n}

\newcommand{\ie}{\textit{i.e.,~}}
\newcommand{\eg}{\textit{e.g.,~}}

\usepackage[inline]{enumitem}
\setlist[description]{leftmargin=\parindent,labelindent=\parindent, font=\normalfont\itshape}

\usepackage{multirow}
\usepackage{tabularx}
\usepackage{cleveref} 

\newcommand{\concat}{\mathop{\Vert}\displaylimits}
\crefname{section}{\S}{\S} 
\crefname{subsection}{\S}{\S} 

\AtBeginDocument{%
  \providecommand\BibTeX{{%
    \normalfont B\kern-0.5em{\scshape i\kern-0.25em b}\kern-0.8em\TeX}}}

\copyrightyear{2026}
\acmYear{2026}
\setcopyright{cc}
\setcctype{by}
\acmConference[SIGIR '26]{Proceedings of the 49th International ACM SIGIR Conference on Research and Development in Information Retrieval}{July 20--24, 2026}{Melbourne, VIC, Australia}
\acmBooktitle{Proceedings of the 49th International ACM SIGIR Conference on Research and Development in Information Retrieval (SIGIR '26), July 20--24, 2026, Melbourne, VIC, Australia}
\acmDOI{10.1145/3805712.3809684}
\acmISBN{979-8-4007-2599-9/2026/07}

\begin{document}

\title[Multi-modal Item Representation Learning]{Multi-modal Relational Item Representation Learning for Inferring Substitutable and Complementary Items} 
\newcommand{\name}{{\textsc{\textsc{MMSC}~}}}

\newtheorem{problem}{Problem}
\newcommand{\subscript}[2]{$#1 _ #2$}

\author{Junting Wang}
\email{junting3@illinois.edu}
\affiliation{%
  \institution{University of Illinois Urbana-Champaign}
  \city{Urbana}
  \state{Illinois}
  \country{USA}
}

\author{Chenghuan Guo}
\email{chenghg@amazon.com}
\affiliation{%
  \institution{Amazon}
  \city{Seattle}
  \state{Washington}
  \country{USA}
}

\author{Yang Jiao}
\email{jaoyan@amazon.com}
\affiliation{%
  \institution{Amazon}
  \city{Seattle}
  \state{Washington}
  \country{USA}
}
\author{Yanhui Guo}
\email{yanhuig@amazon.com}
\affiliation{%
  \institution{Amazon}
  \city{Seattle}
  \state{Washington}
  \country{USA}
}

\author{Hari Sundaram}
\email{hs1@illinois.edu}
\affiliation{%
  \institution{University of Illinois Urbana-Champaign}
  \city{Urbana}
  \state{Illinois}
  \country{USA}
}
\author{Yan Gao}
\email{yanngao@amazon.com}
\affiliation{%
  \institution{Amazon}
  \city{Seattle}
  \state{Washington}
  \country{USA}
}

\setlength{\abovedisplayskip}{0.1cm}
\setlength{\belowdisplayskip}{0.1cm}
\setlength{\floatsep}{0.1cm}
\setlength{\textfloatsep}{0.1cm}
\setlength{\abovecaptionskip}{0.1cm}
\setlength{\belowcaptionskip}{0.1cm}
\setlength{\dbltextfloatsep}{0.1cm}
\setlength{\intextsep}{0.1cm}

\renewcommand{\shortauthors}{Wang et al.}
\newcommand\barbelow[1]{\stackunder[1.2pt]{$#1$}{\rule{.8ex}{.075ex}}}

\makeatletter
\newcommand{\algmargin}{\the\ALG@thistlm}
\makeatother

\setlength{\skip\footins}{5pt}

\begin{abstract}
We study the problem of inferring substitutable and complementary items, which underpins applications such as alternative and follow-up purchase suggestions.
Existing approaches typically learn from behavior-derived item-item associations using GNNs or leverage item content alone.
However, these methods often overlook two key challenges: (i) user behaviors (\eg co-view/co-purchase) only provide noisy weak supervision, and (ii) behavior signals are long-tailed, leaving many items with sparse associations.
We propose \textsc{MMSC}, a self-supervised multi-modal relational representation learning framework that combines a multi-modal foundation model adapted to encode item metadata and a self-supervised denoising module that learns relationship-aware representations from noisy user behaviors, unified by a hierarchical aggregation mechanism. 
We further use LLM-assisted supervision to mitigate noise in behavior-derived supervision during training.
Experiments on five real-world datasets show that \textsc{MMSC} consistently outperforms existing baselines by 26.1\% for substitutable and 39.2\% for complementary item inference, while remaining effective for cold-start items. 
We share our code for reproducibility.\footnote{Available at \url{https://github.com/Junting98/MMSC_code}}

\end{abstract}


\maketitle
\section{Introduction}
\label{sec:introduction}
This paper studies the problem of inferring item-item relationships in e-commerce services, specifically whether two items are \textit{substitutes} or \textit{complements}. These relationships are a core primitive for many downstream services: substitutes enable alternatives for out-of-stock items, while complements support bundle construction and follow-up purchase suggestions. Importantly, our goal is \textbf{not} to learn a user-item recommender; rather, we aim to learn item-item relationships that are independent of specific user preferences.

\begin{figure}[t]
    \centering
    \includegraphics[width=0.9\linewidth]{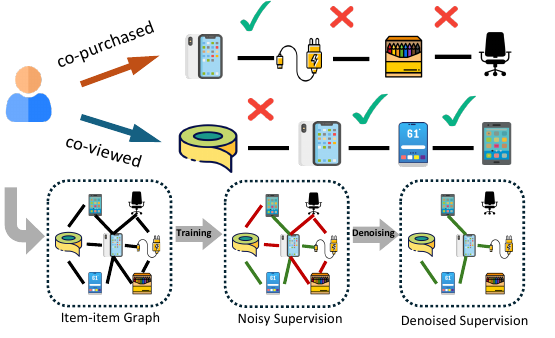}
    \caption{Examples of behavior-derived item pairs and their (in)consistency with true relations. 
\checkmark indicates a plausible relation, while 
$\times$ indicates a spurious pair under the observed behavior. Co-view/co-purchase edges provide weak and noisy supervision for substitutable and complementary inference.}
    \label{fig:intro}

\end{figure}

Modeling substitutable and complementary relationships between items poses two key challenges. \textbf{(i) Noisy weak supervision.} These relationships rarely come with explicit labels and are typically approximated from user behaviors, \eg \textit{co-view} and \textit{co-purchase} edges as weak signals for substitutability and complementarity~\cite{decgcn,dhgan,hetasage,transgat,chen2023enhanced,a2cf,sceptre,ye2023,hao2020p,lva} respectively. However, co-view/co-purchase does not necessarily imply true substitute/complement relations (\Cref{fig:intro}), which introduces label noise during training and also complicates evaluation. \textbf{(ii) Long-tail sparsity.} User behaviors follow heavy-tailed distributions (\Cref{fig:intro_degree}): a small fraction of head items dominates interactions, while most tail items have sparse or missing behavioral links. The combination of noisy weak supervision and long-tail sparsity makes it difficult to infer reliable item-item relationships, especially for cold-start and tail items.

Existing studies on substitutable and complementary inference can be broadly grouped into behavior-only and content-focused methods~\cite{a2cf,dhgan,decgcn,sceptre,ye2023,clva,hetasage}. GNN-based approaches~\cite{decgcn,dhgan,hetasage,transgat,chen2023enhanced} learn item representations from the topology of item-item graphs constructed from behavior co-occurrences, but are sensitive to spurious edges and long-tail sparsity. Other approaches exploit item content~\cite{a2cf,sceptre,ye2023,hao2020p} (\eg with VAEs~\cite{lva}), which is more stable for tail items but often fails to capture fine-grained relational signals implied by user behaviors. Overall, prior work typically treats observed behavior links as reliable and does not explicitly address the joint challenge of noise and sparsity.


\textbf{Our Insight}: User behavior data provides valuable implicit associations between items. However, it is noisy and sparse. In contrast, item metadata (\eg title, description, and images) provides robust descriptors that remain available for tail and cold-start items, yet metadata alone is insufficient to recover relational structure. Therefore, we need to \textit{denoise} behavior-derived links while \textit{integrating} multimodal metadata to obtain representations that are both relationship-aware and robust to sparsity.

\textbf{Present Work}: We propose \textbf{M}ulti-\textbf{M}odal Relational Item Representation Learning for \textbf{S}ubstitutable and \textbf{C}omplementary inference (\textsc{MMSC}), a novel framework that jointly leverages multimodal item metadata and noisy user behaviors to infer substitute and complement relations. Specifically, \name consists of (i) a multi-modal representation module that adapts a multi-modal foundation model to encode item metadata, and (ii) a denoising self-supervised representation module that learns relationship-aware item representations from behavior graphs via contrastive denoising. We further introduce a hierarchical representation aggregation mechanism to combine content- and behavior-based representations.

For model training, inspired by recent advancements in large language models (LLMs)~\cite{bert,gpt3, 10.1145/3774904.3792294,he-etal-2025-llm}, we use LLMs to generate LLM-assisted supervision signals to mitigate the noisy user behaviors. Additionally, we adopt a multi-task learning paradigm to jointly denoise user behaviors and predict substitutable and complementary relationships. \name outperforms existing baselines by 26.1\% for substitutable and 39.2\% for complementary item inference on five real-world datasets. We investigate the effectiveness of each model component through an ablation study. We empirically show that \name also excels in inferring relationships for cold-start items. Our \textbf{key contributions} are as follows:

\begin{figure}[t]
    \centering
    \includegraphics[width=1.15\linewidth,trim=85 0 5 6,clip]{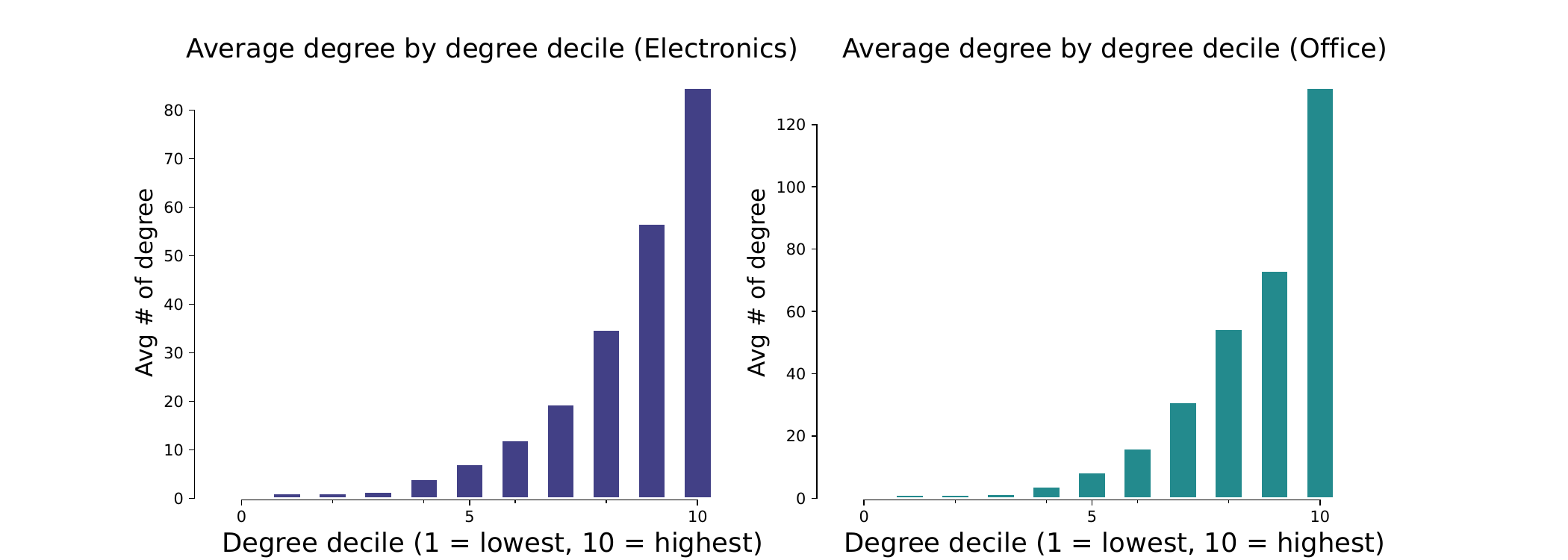}
    \caption{Average degree in the item-item relationship graph. Items are sorted by degree and partitioned into 10 equal-sized groups.}
    \label{fig:intro_degree}
\end{figure}

\begin{description}[labelindent=5pt, labelsep=0pt, topsep=0pt, leftmargin=!, font=\normalfont\bfseries]
\item[Integrative Content-Relational Item Representation:] We propose a framework that jointly leverages item metadata and behavior-derived item-item associations to learn representations for substitutability and complementarity inference.
In contrast to prior work that relies primarily on either content~\cite{a2cf,sceptre,ye2023,hao2020p,lva} or behavior graphs~\cite{decgcn,dhgan,hetasage,transgat}, MMSC fuses metadata-based embeddings (via a multi-modal foundation model with relational adaptation) and behavior-based embeddings (via a meta-path encoder) through hierarchical gating.
Experiments show that combining these two sources consistently improves relation inference, especially under long-tail sparsity.

\item[Noise-aware Learning from Weak Behavior Supervision:] 
We explicitly treat co-view/co-purchase signals as \emph{weak and noisy supervision} and develop a noise-robust training pipeline. Previous methods~\cite{decgcn,dhgan,hetasage,transgat,chen2023enhanced,a2cf,sceptre,ye2023,hao2020p,lva} typically assume reliable behavior data and neglect the noise.
In contrast, we apply self-supervised contrastive learning under structural perturbations to improve robustness to spurious behavior edges and use LLM-assisted validation to refine a subset of behavior-derived positives for cleaner supervision.
Empirically, these components substantially improve substitutability and complementarity inference and are particularly beneficial when behavior signals are noisy.
\end{description}

\section{Related Work}
\label{sec:related_work}

Here, we briefly introduce several lines of related work on substitutable and complementary item inference, as well as multi-modal foundation models.
\subsection{Substitutable and Complementary Relationship Inference}
\label{sec:related_work_sc}
We categorize the related works into two types: behavior-graph (GNN-based) methods~\cite{decgcn,dhgan,hetasage,transgat,chen2023enhanced} and content-based~\cite{a2cf,sceptre,ye2023,hao2020p}.
\subsubsection{\textbf{Behavior-graph (GNN-based) Methods}}

GNN-based methods construct item-item graphs from user behaviors (\eg co-view or co-purchase) and learn item representations by treating items as nodes and behavior signals as edges. 
DecGCN~\cite{decgcn} decouples heterogeneous relations into relation-specific graphs and fuses relation-aware item embeddings via a co-attention mechanism.
DHGAN~\cite{dhgan} extends this line by learning item representations in hyperbolic space.
HetaSAGE~\cite{hetasage} and TransGAT~\cite{transgat} also model behavior graphs with GNN variants, typically focusing on one relation type.
EMRIGCN~\cite{chen2023enhanced} further considers interactions between relation types and proposes a two-level integration scheme to capture shared and relation-specific signals.

Despite strong performance, behavior-graph methods rely on behaviors as weak supervision and thus can be sensitive to noisy edges (\ie behaviors that do not reflect the true relation).
Moreover, they often struggle under cold-start settings where new items have few or no behavioral connections.
Finally, GCN and GAT-based models~\cite{decgcn,dhgan,hetasage,transgat} decouple the original graphs into separate homogeneous graphs, and they ignore valuable connectivity patterns of items (\ie through which relationships are items connected), which may carry useful relational cues.
\subsubsection{\textbf{Content-based Methods}} 
Content-based approaches infer relationships using item metadata, such as textual descriptions, images, or reviews, and are naturally applicable to cold-start items.
Sceptre~\cite{sceptre} learns topic distributions from user reviews using Latent Dirichlet Allocation (LDA~\cite{blei2003latent}), and it uses logistic regression to predict substitutable and complementary relationships.
LVA~\cite{lva} leverages variational auto-encoders (VAEs)~\cite{vae} to encode item content for personalized relationship inference.
A2CF~\cite{a2cf} leverages user reviews to extract item attributes and provide personalized substitutable inference. Ye et al.~\cite{ye2023} use transformers on item textual content to model substitutable relationships. P-companion~\cite{hao2020p} uses an encoder-decoder network to predict multiple complementary item types. 

Compared to behavior-graph methods, content-based models are more robust to missing or sparse behaviors and can handle cold-start items.
However, relying on content alone may miss higher-order relational structures that are informative for real-world substitutable and complementary relations.
\subsection{Multi-modal Foundation Models}
\label{sec:related_work_mm}
Multi-modal foundation models learn aligned representations from paired image-text data and enable strong zero-shot or low-shot transfer across vision-language tasks~\cite{clip,blip,li2023blip2bootstrappinglanguageimagepretraining}.
Recent surveys and works have explored their use in recommender systems by encoding item images and text into transferable item embeddings~\cite{liu2024multimodalrecommendersystemssurvey,10.1145/3589334.3645626,10.1145/3583780.3615039,10.1145/3690624.3709280, li2025llm, preprec, protocf, jiao2024rethinking, liu2025learning}.
However, to the best of our knowledge, prior work has not systematically studied how to adapt multi-modal foundation models to \emph{typed} item-item relationship inference under noisy behavior supervision.



\begin{figure*}[t]
    \centering
    \includegraphics[width=\linewidth]{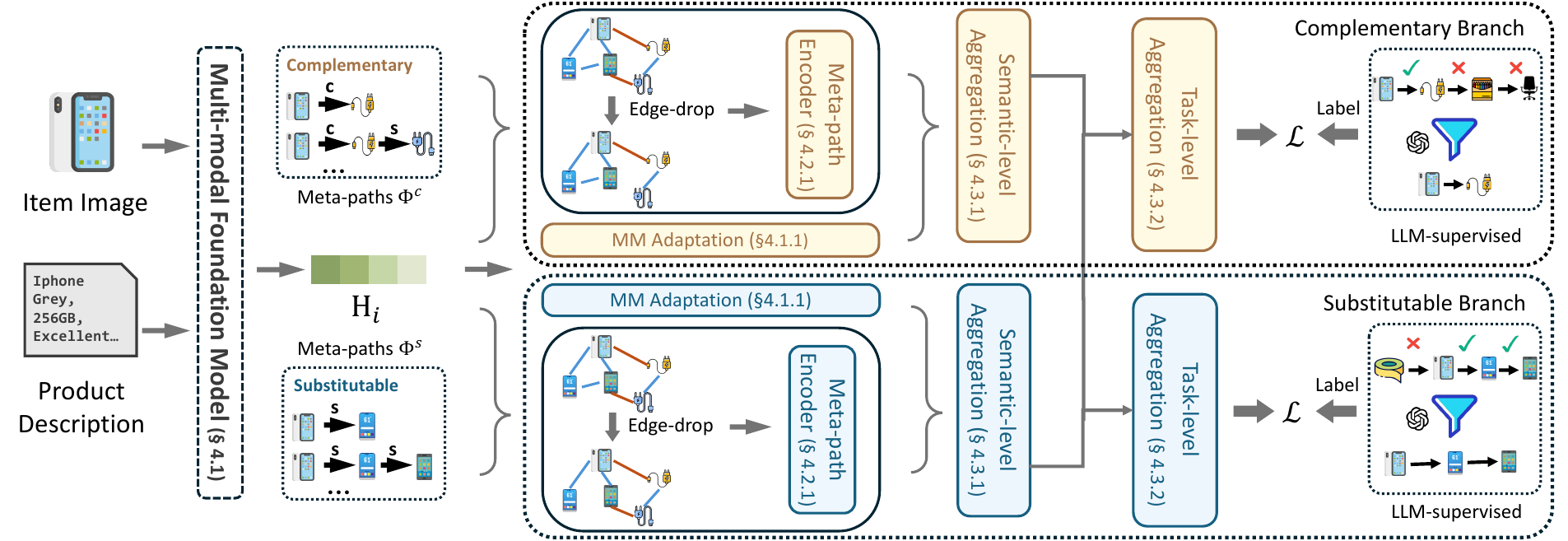}
    \caption{The model architecture of MMSC.}
    \label{fig:model}
\end{figure*}
\section{Preliminaries}
\label{sec:problem_formulation}
\vspace{-2pt}
In this section, we formally define the task of substitutable and complementary item inference and introduce the notation.

\textbf{Substitutable and Complementary Item Inference}: Let $\mathcal{V}=\{v_1,\dots,v_N\}$ denote the set of items, where $N$ is the number of items. Each item $v_i$ is associated with multimodal metadata $x_i$ (\eg text and image), and we denote the collection of metadata as $\mathcal{X}=\{X_i\}_{i=1}^{N}$. We consider two types of item-item relations: substitutable edges $\mathcal{E}^s$ and complementary edges $\mathcal{E}^c$.
This defines a typed item-item graph $\mathcal{G}=(\mathcal{V},\mathcal{E}^s,\mathcal{E}^c)$.\footnote{We treat relations as undirected in this work.} Given a query item $v_i$, the goal is to learn scoring functions
$\mathcal{F}^s(v_j \mid v_i)$ and $\mathcal{F}^c(v_j \mid v_i)$ that score candidate items $v_j \in \mathcal{V}\setminus\{v_i\}$,
such that true substitute/complement items are ranked ahead of non-related items.
Equivalently, we aim to predict whether a typed edge exists between $(v_i,v_j)$:
$y_{ij}^s = \mathbb{I}\{(v_i,v_j)\in\mathcal{E}^s\}$ and
$y_{ij}^c = \mathbb{I}\{(v_i,v_j)\in\mathcal{E}^c\}$.

This is an item-item link prediction problem rather than user-item personalized recommendation.

\section{Methodology}
\label{sec:methods}
We present \textsc{MMSC}, which integrates (i) a metadata-driven multi-modal encoder (\Cref{sec:multi_modal}),
(ii) self-supervised denoising on behavior graphs to reduce noise in behavior supervision (\Cref{sec:behavior_based}),
and (iii) a hierarchical aggregation module that fuses content and behavior signals (\Cref{sec:embedding_aggregation}).
We then describe LLM-assisted supervision and the multi-task training objective in \Cref{sec:multi_task}.
\subsection{\textbf{Multi-modal Item Representation Learning}}
\label{sec:multi_modal}
Multi-modal item metadata (\eg title/description and images) provides robust semantic cues that remain available for tail and cold-start items for modeling item-item relationships. For instance, substitutable items often share similar images or descriptions. Recent advancements in multi-modal foundation models~\cite{li2023blip2bootstrappinglanguageimagepretraining,clip} enable leveraging such content to learn more informative item representations. However, foundation embeddings are not trained to model typed item-item relations (substitute vs.\ complement), and a naive use of their embedding similarity is often suboptimal. We therefore introduce a lightweight \textbf{relational adaptation layer} on top of a frozen multi-modal backbone, producing task-specific representations for substitute and complement inference.

Let $X_i$ denote the multi-modal metadata of item $v_i$, and let $\mathcal{M}$ be a pre-trained multi-modal foundation model (\eg CLIP~\cite{clip} or BLIP-2~\cite{li2023blip2bootstrappinglanguageimagepretraining}) whose parameters are kept \emph{frozen}.
We denote its token-level output as $\mH_i = \mathcal{M}(X_i)\in\mathbb{R}^{m\times d}$, where $m$ is the token length and $d$ is the hidden dimension.

\subsubsection{\textbf{Relational Adaptation via Multi-head Attention.}}
The frozen backbone $\mathcal{M}$ produces base token-level metadata representation $\mH_i$, which captures generic semantics but are not optimized for predicting \emph{typed} item-item relations.
We therefore apply a lightweight \emph{relational adaptation layer}.
Concretely, we use a multi-head attention layer followed by pooling:
\begin{equation}
\label{eq:rel_adapter}
\begin{split}
    \widetilde{\mH}_i &= \text{MHAttn}(\mH_i),\\
    \vq_i &= \text{Pool}(\widetilde{\mH}_i)\in\mathbb{R}^{d},
\end{split}
\end{equation}
where $\text{Pool}(\cdot)$ is mean pooling over tokens.
For MHAttn with $L$ heads and per-head dimension $d_h=d/L$, we use:
\begin{equation}
\label{eq:mhattn}
\begin{split}
    \text{MHAttn}(\mH) &= \text{Concat}(\text{head}_1,\ldots,\text{head}_L)\mW^O,\\
    \text{head}_l &= \text{Attn}(\mH\mW^Q_l,\mH\mW^K_l,\mH\mW^V_l),\\
    \text{Attn}(\mQ,\mK,\mV) &= \text{softmax}\!\left(\frac{\mQ\mK^\top}{\sqrt{d_h}}\right)\mV,
\end{split}
\end{equation}
where $\mW^Q_l,\mW^K_l,\mW^V_l\in\mathbb{R}^{d\times d_h}$ and $\mW^O\in\mathbb{R}^{d\times d}$ are learnable model parameters. 

\subsubsection{\textbf{Task-specific Decoupling.}}
Substitute and complement relations rely on different cues (semantic similarity vs.\ compatibility). Following prior work~\cite{decgcn,dhgan}, to avoid forcing a single embedding to explain both relation types, we decouple the metadata-based representations using separate adaptation layers:
\begin{equation}
\label{eq:q_sc}
\vq_i^s=\text{Pool}(\text{MHAttn}^s(\mH_i)),\quad
\vq_i^c=\text{Pool}(\text{MHAttn}^c(\mH_i)),
\end{equation}
and denote the metadata-driven output as $\vq_i=\{\vq_i^s,\vq_i^c\}$.

\subsection{\textbf{Self-supervised Behavior-based Item Representation Learning}}
\label{sec:behavior_based}
User behaviors (\eg co-view and co-purchase) provide relational signals that connect items and are widely used as weak supervision for substitutable and complementary relationship inference. 
Existing approaches construct behavior-derived item-item graphs and apply GNNs to exploit their topology~\cite{decgcn,dhgan,hetasage,transgat}. 
However, behavior-derived edges are often noisy proxies for true relations (\Cref{fig:intro}) and exhibit long-tail sparsity (\Cref{fig:intro_degree}).
To handle these challenges, we propose a self-supervised behavior-based learning paradigm with two complementary components:
a meta-path encoder that captures higher-order typed association patterns, and
a contrastive denoising objective that enforces representation consistency under structural perturbations, which emulates noisy edges.
\subsubsection{\textbf{User Behavior Encoder via Meta-paths}}
\label{sec:meta-path}
Prior works~\cite{decgcn,dhgan} typically construct separate item-item graphs for different behavior types (\eg co-view for substitutes and co-purchase for complements) and apply GNNs on each graph independently.
Such decoupling can discard higher-order connective patterns in the underlying heterogeneous graph.
 For instance, if $v_1$ is linked to $v_2$ via a substitute edge ($s$) and $v_3$ is linked to $v_2$ via a complement edge ($c$), then the pair $(v_1,v_3)$ may exhibit a meaningful typed association (\eg complementarity) mediated by $v_2$.
  However, this transitive relationship is lost when decoupling the item-item graph, \ie item $v_1$ and $v_3$ will not be directly connected in the decoupled graphs.

\paragraph{\textbf{Meta-path Construction and Neighborhoods.}} We therefore model behavior signals as a typed item-item graph and learn item representations along meta-paths for inferring substitutable and complementary relationships.
 \textit{A meta-path defines a structured pathway connecting nodes of specific types through specific relations in a heterogeneous graph}. Let $s$ be a substitutable relationship and $c$ be a complementary relationship. A meta-path $v_1 \xrightarrow{s} v_2 \xrightarrow{c} v_3 \xrightarrow{s} v_4$ connects $v_1$ and $v_4$ through a structured composition of relation types, enabling the encoder to capture fine-grained relational patterns beyond direct edges. In this case, the meta-path may indicate higher-order patterns useful for predicting complementarity ($v_1$ and $v_4$).

We define two relation-specific meta-path sets: $\Phi^s$ for substitutable and $\Phi^c$ for complementary inference respectively.
Let $\Phi^r=\{\phi_1^r,\ldots,\phi_{K_r}^r\}$ denote the meta-path set for relation $r\in\{s,c\}$, where $K_r$ is the number of meta-paths.
For an item $v_i$ and a meta-path $\phi\in\Phi^r$, we denote by $\mathcal{N}_i^{\phi}$ the set of neighbors reachable from $v_i$ following $\phi$.
To aggregate neighborhood information, we employ a node-level GAT-style neighbor attention and a path-level attention over meta-paths.

\paragraph{\textbf{Node-level Attention.}}
For a relation type $r\in\{s,c\}$ and a meta-path $\phi\in\Phi^r$, the node-level attention aggregates the meta-path neighbors $\mathcal{N}^{\phi}_i$ of item $v_i$ to obtain a meta-path-specific representation $\vz^{\phi}_i$. 
Specifically, we first compute the attention score between $v_i$ and its neighbors $v_j\in\mathcal{N}^{\phi}_i$ as follows:
\begin{equation}
    \alpha^{\phi}_{ij} = 
    \frac{\exp\!\left(\text{LeakyReLU}\!\left(\mW^\intercal_{\phi}[\vh_i||\vh_j]\right)\right)}
    {\sum_{v_t \in \mathcal{N}^{\phi}_i} \exp\!\left(\text{LeakyReLU}\!\left(\mW^\intercal_{\phi}[\vh_i||\vh_t]\right)\right)},
\end{equation}
where $\mW_{\phi} \in \mathbb{R}^{2d \times 1}$ is the weight vector for meta-path $\phi$, $||$ denotes concatenation, $\cdot^\intercal$ denotes transposition, and $\text{LeakyReLU}$~\cite{leakyrelu} is the nonlinear activation.
We use the output of the multi-modal foundation model $\vh_i = \text{MeanPool}(\mH_i) \in \mathbb{R}^d$ as the input node feature, which injects multi-modal semantics into behavior-based relational learning.
The node-level aggregation is then computed as:
\begin{equation}
    \vz^{\phi'}_i = \sigma\Bigl(\sum_{v_j \in \mathcal{N}^{\phi}_i} \alpha^{\phi}_{ij} \mW_a^\intercal \vh_j\Bigr),
\end{equation}
where $\mW_a \in \mathbb{R}^{d \times d}$ is a learnable weight matrix and $\sigma$ is the ELU activation~\cite{elu}.
To attend to information from different representation subspaces and stabilize training, we adopt a multi-head variant of the above \emph{neighbor attention}~\cite{han,gat,vaswani2017attention}.
Specifically, we repeat the node-level attention $T$ times with $T$ independent attention mechanisms and concatenate the learned embeddings:
\begin{equation}
    \vz^{\phi}_i =  
    \concat^{T}_{t=1} 
    \sigma\Bigl(\sum_{v_j \in \mathcal{N}^{\phi}_i} \alpha^{\phi,t}_{ij} (\mW_a^t)^\intercal \vh_j\Bigr),
\end{equation}
where $\alpha^{\phi,t}_{ij}$ is the attention coefficient of the $t$-th head and $\mW_a^t$ is its corresponding projection matrix.
\paragraph{\textbf{Path-level Attention.}}
Different meta-paths capture different fine-grained typed associations between items; thus, their meta-path-specific representations $\{\vz^{\phi}_i\}$ are complementary to each other.
To learn a more informative representation of item $v_i$, we aggregate the meta-path representations via a path-level attention mechanism to obtain the semantic-level representation $\vp_i^{r}$.
We define the importance of each meta-path embedding as:
\begin{equation}
    \begin{split}
        w^{\phi}_i &=  \vs^\intercal  \tanh(\mW_b \cdot \vz^{\phi}_i + \vb), \\
        \beta^{\phi}_i &= \frac{\exp(w^{\phi}_i)}{\sum_{\phi' \in \Phi^r} \exp(w^{\phi'}_i)},
    \end{split}
\end{equation}
where $\mW_b \in \mathbb{R}^{d \times d}$, $\vb \in \mathbb{R}^{d}$, and $\vs \in \mathbb{R}^{d}$ are learnable parameters, and $\tanh(\cdot)$ is the hyperbolic tangent activation.
The final semantic-level representation is computed as:
\begin{equation}
    \vp_i^{r} = \sum_{\phi \in \Phi^r} \beta^{\phi}_i \vz^{\phi}_i.
\end{equation}
Similar to the multi-modal branch, we decouple behavior-based representations for the two relation types.
We compute $\vp_i^s$ and $\vp_i^c$ using relation-specific meta-path sets $\Phi^s$ and $\Phi^c$ (see \Cref{sec:implementation} for a detailed list of the meta-paths), with separate encoders for each set.
The behavior-based output is $\vp_i=\{\vp_i^s,\vp_i^c\}$.
\subsubsection{\textbf{Self-supervised Behavior Denoising}}
To learn robust representations under noisy behavior supervision, we enforce invariance to structural perturbations that simulate potential noise in item-item associations.
To this end, we propose a self-supervised objective for denoising and robustness enhancement.
We utilize edge dropout~\cite{wu2021self} by randomly removing a fraction of behavior edges to form a perturbed view $\mathcal{G}'$.
We denote the perturbed graph as $\mathcal{G}' = (\mathcal{V}, \mathcal{E}'^{s}, \mathcal{E}'^{c})$.
Applying the same meta-path encoder yields alternative embeddings $\vp_i'=\{\vp_i^{s'},\vp_i^{c'}\}$.
  
We treat two views of the same item as a positive pair (\eg $\vp_i^s$ and $\vp_i^{s'}$) and views of different items as negatives. To enhance the robustness of item representations, we minimize the contrastive loss between the positive and negative pairs. Specifically, we optimize an InfoNCE~\cite{infonce} objective with in-batch negatives:
\begin{equation}
    \begin{split}
        \mathcal{L}_{\text{self}}^s &= - \frac{1}{|\mathcal{V}|}\sum_{v_i \in \mathcal{V}} \log \frac{\exp(s(\vp_i^s, \vp_i^{s'})/\tau)}{\sum_{j \neq i} \exp(s(\vp_i^s, \vp_j^s)/\tau)} \\
        \mathcal{L}_{\text{self}}^c &= - \frac{1}{|\mathcal{V}|}\sum_{v_i \in \mathcal{V}} \log \frac{\exp(s(\vp_i^c, \vp_i^{c'})/\tau)}{\sum_{j \neq i} \exp(s(\vp_i^c, \vp_j^c)/\tau)} \\
        \mathcal{L}_{\text{self}} &= \mathcal{L}_{\text{self}}^s + \mathcal{L}_{\text{self}}^c
    \end{split}
    \label{eqn:self_supervised}
\end{equation}
    where $s$ is the cosine similarity function, and $\tau$ is the temperature parameter. We will discuss the optimization in~\Cref{sec:multi_task}. We use InfoNCE to enforce consistency between original and perturbed graph views while keeping different item representations distinguishable through in-batch negatives, matching our goal of denoising behavior-based representations without additional labels.

\subsection{\textbf{Hierarchical Representation Aggregation}}
We propose a hierarchical aggregation strategy that fuses metadata-driven and behavior-driven item representations at the \emph{semantic} level and substitutability- and complementarity-specific representations at the \emph{task} level.

\label{sec:embedding_aggregation}
\subsubsection{\textbf{Semantic-Level Aggregation}}
The metadata-based representation $\vq_i$ (from \Cref{sec:multi_modal}) captures item semantics and is available for all items, while the behavior-based representation $\vp_i$ (from \Cref{sec:behavior_based}) encodes high-order relational patterns but can be sparse/noisy. Aggregating these two representations ensures that the aggregated representation effectively integrates both aspects, which are essential for modeling substitutable and complementary relationships.

We combine them via a gating mechanism that adaptively balances the two sources per feature dimension.
 Specifically, we use a neural gating mechanism that learns a non-linear gate, $\vg$, to control the flow of information between these representations. Without loss of generality, we demonstrate the gating mechanism for learning the item representation $\va^{s}_i$ for the substitutability inference:
\begin{equation}
\begin{split}
    \va^{s}_i &= \text{Gating}_{\text{sem}}^s(\vq^s_i, \vp^s_i)\\
     \text{Gating}_{\text{sem}}^s(\vq^s_i, \vp^s_i)&=\vg \odot \vp^s_i + (1-\vg) \odot \vq^s_i, \\
     \vg &= \sigma(\mW_{g_1}\vp^s_i +\mW_{g_2}\vq^s_i + \vb_g)\\
\end{split}
\label{eqn:gate}
\end{equation}
where $\sigma$ is the sigmoid function, $\odot$ denotes the element-wise multiplication, $\mW_{g_1}\in \mathbb{R}^{d\times d}$, $\mW_{g_2}\in \mathbb{R}^{d\times d}$, and $\vb_g\in \mathbb{R}^{d}$ are learnable parameters. 
The complementary semantic-level representation $\va^{c}_i$ is computed analogously using $\text{Gating}_{\text{sem}}^c(\vq^c_i,\vp^c_i)$ with its own parameters.
\vspace{-5pt}
\subsubsection{\textbf{Task-Level Aggregation}}
\label{sec:task_level}
Although substitutability and complementarity represent different relation types, their representations can share useful signals (\eg overlapping semantics and shared contexts)~\cite{dhgan,decgcn}. 
We therefore allow controlled cross-task information sharing by applying another gating mechanism over $\va_i^{s}$ and $\va_i^{c}$. Similarly, without loss of generality, we demonstrate the gating mechanism for learning the final item representation for the substitutability inference:
\begin{equation}
\begin{split}
    \ve^{s}_i &= \text{Gating}_{\text{task}}^s(\va^{s}_i, \va^{c}_i)
              = \gamma^{s}_i \odot \va^{s}_i + (1-\gamma^{s}_i)\odot \va^{c}_i, \\
    \gamma^{s}_i &= \sigma(\mW_{t_1}^s \va^{s}_i + \mW_{t_2}^s \va^{c}_i + \vb_t^s),\\
\end{split}
\label{eqn:gate_task}
\end{equation}
where $\mW_{t_1}^s\in \mathbb{R}^{d\times d}$, $\mW_{t_2}^s\in \mathbb{R}^{d\times d}$, and $\vb_t^s\in \mathbb{R}^{d}$ are learnable parameters. The complementary representation $\ve^{c}_i$ is computed analogously using $\text{Gating}_{\text{task}}^c(\va^{s}_i,\va^{c}_i)$ with its own parameters.
The final representation of item $v_i$ is $\ve_i=\{\ve_i^s,\ve_i^c\}$.
\newcommand*{\factor}{0.10}
\begin{table}[t]
    \centering 
    \small 
    \setlength{\tabcolsep}{0pt} 
    \renewcommand{\arraystretch}{0.3} 
    \begin{tabular}{p{\linewidth}}
    \toprule
    \multicolumn{1}{l}{\textbf{Template of LLM-assisted Supervision for Substitutable Items:}} \\
    \midrule
    \cellcolor{gray!10}{
    Answer the following question with yes or no only.  I am considering two items "$\{$asin\_x$\}$" and "$\{$asin\_y$\}$".  If one of them is out-of-stock, can I buy the other one to serve the same purpose?}  \\
    
    \midrule
    \multicolumn{1}{l}{\textbf{Template of LLM-assisted Supervision for Complementary Items:}} \\
    \midrule
    \cellcolor{gray!10}{
    Here is one example of two items that if I bought one item then I can also buy the other to serve as a complementary:
    The two items are Sheaffer(R) Pen Refills, Ink Cartridges, Jet Black, Pack Of 5 and Sheaffer 100 Red Fountain Pen 9307-0.

    Answer the following question with yes or no only.
    I am considering two items \{asin\_x\} and \{asin\_y\}, 
    if I bought one item, then can I buy the other to serve as a complementary?}  \\
    \bottomrule
    \end{tabular}
    \caption{Prompts for filtering user behaviors (\Cref{sec:llm_augmented}). We replace $\{$asin\_x$\}$ and $\{$asin\_y$\}$ with the metadata (\eg description and title) of the items with the corresponding ASINs.}
    \label{tab:llm_template}
\vspace{-7pt}
    \end{table}

\vspace{-5pt}
\subsection{LLM-Assisted Supervision and Multi-Task Learning}
\label{sec:multi_task}
We use LLMs to refine behavior-derived weak supervision and jointly optimize substitutability and complementarity inference together with the self-supervised objective.
\subsubsection{\textbf{LLM-Assisted Supervision}}
\label{sec:llm_augmented}
Prior works~\cite{decgcn,dhgan,yan2022personalized,hao2020p} derive training labels from user behaviors, using \textit{co-view} as a proxy for substitutability and \textit{co-purchase} as a proxy for complementarity. 
However, these behavior-derived edges are noisy proxies for the true relations (\eg co-purchase can contain unrelated or even substitutable items; \Cref{fig:intro}).
To mitigate this noise, we employ LLMs with prompts (\Cref{tab:llm_template}) to validate item pairs and construct a cleaner supervision set.
Specifically, we sample a subset of behavior-derived item pairs and query an LLM to judge whether each pair is substitutable or complementary.
The resulting LLM-validated edges form $\mathcal{E}_{\text{LLM}}=\mathcal{E}_{\text{LLM}}^s\cup \mathcal{E}_{\text{LLM}}^c$, which provides more reliable supervision (\Cref{sec:dataset_analysis}).

While LLMs can assess item-item relations effectively, their inference cost makes them impractical for large-scale scoring over all item pairs.
Therefore, we only use $\mathcal{E}_{\text{LLM}}$ as \emph{training supervision}, while learning item representations for efficient inference at scale.
\subsubsection{\textbf{Multi-task Learning}}
\label{sec:multi_task_learning}
We jointly optimize the supervised objectives for substitutability and complementarity inference together with the self-supervised objective in \Cref{eqn:self_supervised}.
For supervised learning, we adopt a triplet ranking loss with LLM-validated edges as positives:
\begin{equation}
    \begin{split}
        \mathcal{L}_{\text{triplet}}^s &= \sum_{(v_i, v_j^+) \in \mathcal{E}_{\text{LLM}}^s}
        \max\bigl(0, m + s(\ve_i^s, \ve_{j^+}^{s}) - s(\ve_i^s, \ve_{k^-}^{s})\bigr),\\
        \mathcal{L}_{\text{triplet}}^c &= \sum_{(v_i, v_j^+) \in \mathcal{E}_{\text{LLM}}^c}
        \max\bigl(0, m + s(\ve_i^c, \ve_{j^+}^{c}) - s(\ve_i^c, \ve_{k^-}^{c})\bigr),\\
        \mathcal{L}_{\text{triplet}} &= \mathcal{L}_{\text{triplet}}^s + \mathcal{L}_{\text{triplet}}^c,
    \end{split}
\end{equation}
\label{eq:triplet_loss}
where $s(\cdot,\cdot)$ denotes cosine similarity, $m$ is the margin, and $v_k^-$ is a negative item sampled for anchor $v_i$ (\eg from behavior-linked candidates that are not in $\mathcal{E}_{\text{LLM}}^r$). We use a triplet ranking loss because the task is evaluated by ranking true related items above unrelated candidates. The margin directly encourages separation between LLM-validated positives and sampled negatives under noisy weak supervision.
\renewcommand*{\factor}{0.10}

\begin{table}[t]
    \centering
    \small
    \begin{tabular}{@{}lccccc@{}}
    \toprule
    \textsc{Dataset}  &  {\textsc{\#items}} & {\textsc{\#co-view} $\cup $ \textsc{\#buy-after-view}}& {\textsc{\#co-purchase}}\\ 
    \midrule
    Office  &85,648&2,002,109&1,200,487\\
    Tool  &192,408&3,413,564&2,058,000\\ 
    Toys  &227,227&8,718,700&6,784,468\\
    Home  &250,244&7,297,466&2,070,856\\
    Electronics  &326,002&6,120,988&3,816,044\\
    \bottomrule
    \end{tabular}
    \caption{Dataset statistics.}
    \label{tab:dataset_statistics}
    \end{table}

Finally, the overall multi-task learning objective is:
\begin{equation}
    \mathcal{L} = \mathcal{L}_{\text{triplet}} + \lambda \mathcal{L}_{\text{self}}
\end{equation}
\label{eq:final_loss}
where $\lambda$ is the hyperparameter that controls the weight of the self-supervised objective.


\section{Experiments}
\label{sec:experiments}
We conduct experiments on five real-world datasets and address the following research questions:
\begin{enumerate*}[label=\textbf{RQ\arabic*:} ]
\item How does \name perform compared to state-of-the-art methods for inferring substitutable and complementary items?
\item How do different components of \name contribute to performance?
\item How effective is \name for cold-start items with no observed behavior edges?
\item How does \name perform under noisy and sparse behavior supervision?
\item How sensitive is \name to key hyperparameters?
\end{enumerate*}
\subsection{Datasets and Experimental Details}
\label{subsec:datasets}

\subsubsection{\textbf{Datasets and Preprocessing}}
We conduct experiments on the Amazon review dataset~\cite{mcauley2015image}, following prior work~\cite{decgcn,dhgan,clva}.
We use five categories: Office Products, Tools and Home Improvement, Electronics, Toys and Games, and Home and Kitchen.
Dataset statistics are reported in~\Cref{tab:dataset_statistics}.

To ensure the quality of items, we filter out items with missing textual metadata or images.
For each item, we use its title and description as text and its product image as visual input.
Following prior work~\cite{decgcn,dhgan,hao2020p}, we formulate substitutable and complementary relationship inference as a link prediction task on a behavior-derived item-item graph:
we treat \textit{co-view} and \textit{buy-after-view} as weak signals for substitutability, and \textit{co-purchase} as a weak signal for complementarity.

\paragraph{\textbf{Train/Test Split.}}
For each item that has at least one neighbor under a given relation type, we randomly sample one incident edge as a test candidate for that relation type and use the remaining edges for training. Items without any edge of a relation type do not contribute test edges for that type.
We remove all sampled test edges from the training graph to avoid leakage.

\paragraph{\textbf{LLM-filtered Evaluation Sets.}}
Because behavior-derived edges are noisy, we refine the test candidates using an LLM with prompts in~\Cref{tab:llm_template} to obtain higher-precision evaluation sets.
The refined test sets are denoted as $Y_{\text{sub}}$ and $Y_{\text{com}}$ for substitutable and complementary relations, respectively.
The LLM is used only to construct evaluation/training supervision; it is \emph{not} used at inference time.

\subsubsection{\textbf{Quantifying Noise in Weak Supervision.}}
\label{sec:dataset_analysis}
To empirically validate that behavior-derived item pairs provide noisy weak supervision, we conduct a case study on 100 randomly sampled item pairs from the Office dataset. We compare behavior-derived labels and LLM-derived labels against human annotations. Behavior-derived labels achieve 78.0\% accuracy for substitutability and 24.5\% for complementarity, indicating substantial noise in co-view/co-purchase signals, especially for complementary relations. In contrast, LLM-derived labels achieve 94.7\% and 57.9\% accuracy, respectively, suggesting that LLM filtering provides more reliable supervision for both relation types.

\begin{table}[t]
    \centering
    \small
    \begin{tabular}{llcc}
        \toprule
        \textsc{Dataset} & \textsc{Model} & \textsc{s/epoch, B=128} & \textsc{s/epoch, B=256} \\
        \midrule
        \multirow{4}{*}{\textsc{Office}}
            & DecGCN~\cite{decgcn} & 2.48 & 4.68 \\
            & DHGAN~\cite{dhgan}  & 2.71 & 4.78 \\
            & MMSC   & 0.19* & 0.35* \\
            & GAT~\cite{gat}    & \textbf{0.02} & \textbf{0.03} \\
        \midrule
        \multirow{4}{*}{\textsc{Electronics}}
            & DecGCN~\cite{decgcn} & 2.82 & 5.01 \\
            & DHGAN~\cite{dhgan}  & 3.02 & 5.53 \\
            & MMSC   & 0.20* & 0.36* \\
            & GAT~\cite{gat}    & \textbf{0.03} &\textbf{0.04} \\
        \bottomrule
    \end{tabular}
    \caption{Training time per epoch (seconds) for different batch sizes $B$ across datasets. Larger number means longer training time.}
    \label{tab:training_time}
\end{table}
    \newcommand*{\mainfactor}{0.038}
\begin{table*}[t]
    \centering
    \small 
    \begin{tabular}{p{0.11\linewidth}p{\mainfactor\linewidth}p{\mainfactor\linewidth}p{\mainfactor\linewidth}p{\mainfactor\linewidth}p{\mainfactor\linewidth}p{\mainfactor\linewidth}p{\mainfactor\linewidth}p{\mainfactor\linewidth}p{\mainfactor\linewidth}p{\mainfactor\linewidth}p{\mainfactor\linewidth}p{\mainfactor\linewidth}p{\mainfactor\linewidth}p{\mainfactor\linewidth}p{\mainfactor\linewidth}}
        {\textsc{Dataset}} & \multicolumn{3}{c}{\textsc{Office}} & \multicolumn{3}{c}{\textsc{Tools}} & \multicolumn{3}{c}{\textsc{Toys}}&\multicolumn{3}{c}{\textsc{Home}} & \multicolumn{3}{c}{\textsc{Electronics}}\\
    \toprule
    \textsc{Substitutable} &H@10&M@10&N@10 &H@10&M@10&N@10 &H@10&M@10&N@10 &H@10&M@10&N@10 &H@10&M@10&N@10\\
    \midrule
    \multicolumn{16}{c}{\textsc{Graph Neural Networks}}\\
    GATNE-I~\cite{gatne} & 0.629&0.384&0.443&0.634&0.367&0.430&0.659&0.435&0.489&0.555&0.324&0.379&0.611&0.348&0.410\\
    GAT~\cite{gat}&0.758&0.490&0.554&0.755&0.494&0.557&0.764&0.511&0.572& 0.728&0.455&0.520&0.728&0.466&0.528\\
    HAN~\cite{han}&0.796&0.500&0.571&0.765&0.460&0.532&0.759&0.492&0.556& 0.779&0.500&0.567&0.775&0.469&0.542\\
    \multicolumn{16}{c}{\textsc{Substitute and Complement Relation Models}}\\
    DecGCN~\cite{decgcn}&0.561&0.302&0.363&0.637&0.344&0.413&0.573&0.318&0.378&0.533&0.265&0.328&0.617&0.318&0.396\\
    DHGAN~\cite{dhgan}&0.866*&0.573&0.644*&0.916*&0.652&0.716*&0.901*&0.658*&0.717*&0.930*&0.727*&0.777*&0.916*&0.652*&0.716*\\
    \multicolumn{16}{c}{\textsc{Adapted Multi-modal Foundation Models}}\\
    BLIP-2~\cite{li2023blip2bootstrappinglanguageimagepretraining} &0.779&0.519&0.581&0.889&0.658*&0.714&0.857&0.587&0.652&0.901&0.692&0.744&0.814&0.547&0.611\\
    CLIP~\cite{clip}&0.841&0.580*&0.642&0.880&0.644&0.701&0.751&0.527&0.583&0.911&0.685&0.740&0.688&0.483&0.532\\
    \midrule
    \name &\textbf{0.980}&\textbf{0.782}&\textbf{0.831}&\textbf{0.989}&\textbf{0.841}&\textbf{0.877}&\textbf{0.978}&\textbf{0.827}&\textbf{0.864}&\textbf{0.984}&\textbf{0.830}&\textbf{0.869}&\textbf{0.989}&\textbf{0.816}&\textbf{0.859}\\

    \% Improvement & $+13.1\%$ &$+36.5\%$&$+29.0\%$&$+7.9\%$&$+29.0\%$&$+22.5\%$&$+8.5\%$&$+25.7\%$&$+20.5\%$&$+5.8\%$&$+14.2\%$&$+11.8\%$&$+8.0\%$&$+25.2\%$&$+20.0\%$ \\

    \midrule
    \midrule
    \textsc{Complementary} &H@10&M@10&N@10 &H@10&M@10&N@10 &H@10&M@10&N@10 &H@10&M@10&N@10 &H@10&M@10&N@10\\
    \midrule

    \multicolumn{16}{c}{\textsc{Graph Neural Networks}}\\
    GATNE-I~\cite{gatne} & 0.690&0.487&0.536&0.780&0.521&0.583&0.786&0.570&0.622&0.807&0.594&0.645&0.781&0.521&0.583\\
    GAT~\cite{gat}&0.769*&0.567*&0.615*&0.844*&0.669*&0.712*&0.844*&0.609*&0.665*&0.875&0.713*&0.752*&0.830*&0.589*&0.647*\\
    HAN~\cite{han}&0.769&0.534&0.590&0.646&0.415&0.470&0.766&0.530&0.587&0.672&0.446&0.500&0.775&0.469&0.542\\
    \multicolumn{16}{c}{\textsc{Substitute and Complement Relation Models}}\\
    DecGCN~\cite{decgcn} &0.630&0.379&0.440&0.727&0.459&0.523&0.573&0.318&0.378&0.573&0.318&0.378&0.620&0.372&0.415\\
    DHGAN~\cite{dhgan}&0.761&0.540&0.529&0.826&0.585&0.643&0.761&0.529&0.529&0.897&0.681&0.733&0.825&0.546&0.612\\
    \multicolumn{16}{c}{\textsc{Adapted Multi-modal Foundation Models}}\\
    BLIP-2~\cite{li2023blip2bootstrappinglanguageimagepretraining} &0.727&0.482&0.541&0.834&0.621&0.672&0.727&0.482&0.541&0.854&0.628&0.682&0.632&0.428&0.476\\
    CLIP~\cite{clip}&0.763&0.553&0.603&0.803&0.618&0.662&0.708&0.493&0.545&0.838&0.644&0.690&0.566&0.405&0.443\\
    \midrule
    \name &\textbf{0.964}&\textbf{0.789}&\textbf{0.832}&\textbf{0.986}&\textbf{0.885}&\textbf{0.911}&\textbf{0.980}&\textbf{0.863}&\textbf{0.892}&\textbf{0.980}&\textbf{0.872}&\textbf{0.899}&\textbf{0.978}&\textbf{0.821}&\textbf{0.860}\\

    \% Improvement &$+25.4\%$&$+39.2\%$&$+35.3\%$&$+16.8\%$&$+32.3\%$&$+28.0\%$&$+16.1\%$&$+62.8\%$&$+34.1\%$&$+9.3\%$&$+22.3\%$&$+19.6\%$&$+17.8\%$&$+39.4\%$&$+32.9\%$ \\
\bottomrule
    \end{tabular}
    \caption{Substitutable and complementary relationship inference results.
Best performance is in bold; second-best is marked with an asterisk (*).
\% improvement is computed relative to the second-best method.
On average, \name improves M@10 by 26.1\% (substitutable) and 39.2\% (complementary) across five datasets.}
    \label{tab:amazon-results}
\end{table*}

\subsubsection{\textbf{Evaluation Protocol}}
For each test edge in $Y_{\text{sub}}$ and $Y_{\text{com}}$, we uniformly sample 1000 negative items that are not connected to the source item under the same relation type.
We rank the true target item against the 1000 negatives and report Hit Ratio (H@10), MRR@10 (denoted as M@10), and NDCG@10 (N@10).

\subsubsection{\textbf{Baselines}}
\label{subsec:baselines}
We compare with baselines from three categories:
(i) graph neural networks (GATNE-I~\cite{gatne}, GAT~\cite{gat}, HAN~\cite{han}),
(ii) substitutable and complementary relation models (DecGCN~\cite{decgcn}, DHGAN~\cite{dhgan}),
and (iii) adapted multi-modal foundation models (CLIP~\cite{clip}, BLIP-2~\cite{li2023blip2bootstrappinglanguageimagepretraining}). For graph-based baselines, we follow the original implementations/settings. DecGCN and DHGAN primarily learn item representations from behavior-derived graph structure with trainable item embeddings / original node features.

To adapt CLIP/BLIP-2 to link prediction, we add a relation-specific MLP on top of the item embeddings.
For fair comparison, we exclude methods without public implementations~\cite{transgat,hetasage,chen2023enhanced,a2cf,sceptre,ye2023,hao2020p,lva}.

\begin{table}[t]
    \centering
    \small
    \begin{tabular}{lccc}
        \toprule
        \textsc{Dataset} & \textsc{Batch Size} & \textsc{Sub. Time (h)} & \textsc{Comp. Time (h)} \\
        \midrule
        \multirow{2}{*}{\textsc{Office}}
        & 8  & 7.92  & 10.08 \\
        & 14 & 6.32  & 9.28  \\
        \midrule
        \multirow{2}{*}{\textsc{Electronics}}
        & 8  & 10.38 & 11.33 \\
        & 14 & 9.33  & 10.53 \\
        \bottomrule
    \end{tabular}
    \caption{Offline LLM preprocessing time for constructing $E_{\mathrm{LLM}}$ on Office and Electronics under different inference batch sizes. Sub. and Comp. denote substitutable and complementary filtering, respectively.}
    \label{tab:llm_preprocessing_time}
    \vspace{-5pt}
\end{table}


\subsubsection{\textbf{Implementation Details}} 
\label{sec:implementation}
We sample five negative items per positive training edge. We use publicly available implementations for the baselines and develop our implementation of \textsc{MMSC} based on the DHGAN~\cite{dhgan} codebase.
  We use Adam optimizer and tune the learning rate in the range $\{10^{-4}, 10^{-3}, 10^{-2}\}$. We set dropout to 0.2 and tune $\lambda$ in ~\Cref{eq:final_loss} in the range $\{10^{-3}, 5^{-3}, 10^{-2}\}$. For every dataset, we fix the size of $\mathcal{E}_{\text{LLM}}$ to be 500K.
We use Flan-T5-XXL~\cite{t5} for LLM filtering and BLIP-2~\cite{li2023blip2bootstrappinglanguageimagepretraining} as the multi-modal foundation model.
We freeze the BLIP-2 parameters and train only the relational adaptation layer (\Cref{sec:multi_modal}).

\paragraph{\textbf{Meta-paths.}}
For the behavior encoder, we explore up to 3-hop neighborhoods.
We use $\Phi^s=\{v_1 \xrightarrow{s} v_2,\ v_1 \xrightarrow{s} v_2 \xrightarrow{s} v_3,\ v_1 \xrightarrow{s} v_2 \xrightarrow{s} v_3 \xrightarrow{s} v_4\}$
and
$\Phi^c=\{v_1 \xrightarrow{c} v_2,\ v_1 \xrightarrow{c} v_2 \xrightarrow{s} v_3,\ v_1 \xrightarrow{s} v_2 \xrightarrow{c} v_3,\ v_1 \xrightarrow{s} v_2 \xrightarrow{s} v_3 \xrightarrow{c} v_4,\ v_1 \xrightarrow{s} v_2 \xrightarrow{c} v_3 \xrightarrow{s} v_4,\ v_1 \xrightarrow{c} v_2 \xrightarrow{s} v_3 \xrightarrow{s} v_4\}$.


\subsubsection{\textbf{Runtime Analysis and LLM Preprocessing Cost}}
We report training time per epoch under different batch sizes in~\Cref{tab:training_time}.
\name is substantially faster than DecGCN and DHGAN while slower than GAT, a lightweight GNN.
All methods are run under the same hardware and software environment, and we report end-to-end epoch time including forward/backward passes and neighborhood sampling. Both the LLM-assisted supervision and the token-level metadata representation are pre-computed and do not contribute to training time.


\paragraph{\textbf{LLM Preprocessing Cost.}}
We use Flan-T5-XXL~\cite{t5} for LLM-assisted supervision, which has 11B parameters. \Cref{tab:llm_preprocessing_time} reports the offline preprocessing time for generating LLM-assisted training supervision on a single RTX 4090 GPU. This preprocessing is a one-time offline cost and is not incurred during model inference or online serving. The cost can be further reduced by parallelizing candidate-pair filtering across multiple GPUs/machines or by increasing the inference batch size when memory allows.

\renewcommand*{\factor}{0.065}
\begin{table}[t]
    \centering
    \small
    \begin{tabular}{p{0.3\linewidth}p{\factor\linewidth}p{\factor\linewidth}p{\factor\linewidth}p{\factor\linewidth}p{\factor\linewidth}
        p{\factor\linewidth}}
         & \multicolumn{3}{c}{\textsc{Substitutable}}& \multicolumn{3}{c}{\textsc{Complementary}}\\
    \toprule
    {\textsc{Office}}&H@10&M@10&N@10&H@10&M@10&N@10\\
    \midrule
    MMSC &\textbf{0.980} &\textbf{0.782} &\textbf{0.831} &\textbf{0.964} &\textbf{0.789} &\textbf{0.832} \\
    w/o SSL (\Cref{eqn:self_supervised}) &0.968&0.749&0.803&0.958&0.757&0.801\\
    w/o TA (\Cref{sec:task_level})&\textbf{0.980}&0.776*&0.827*&0.965*&0.775*&0.822*\\
    w/o SSL \& TA &0.976*&0.757&0.811&0.867&0.649&0.731\\
    w/o MM (\Cref{sec:multi_modal})&0.911&0.603&0.678&0.837&0.594&0.660\\
    w/o BM (\Cref{sec:behavior_based})&0.849&0.560&0.630&0.727&0.482&0.541\\
    w/o $\mathcal{E}_{\text{LLM}}$ (\Cref{sec:multi_task})&0.967&0.727&0.786&0.940&0.718&0.772\\
    w/o 3rd-hop (\Cref{sec:implementation}) &0.972&0.750&0.804&0.953&0.730&0.785\\
    w/o 3rd-hop, $\mathcal{E}_{\text{LLM}}$&0.968&0.739&0.795&0.949&0.751&0.799\\
   \toprule
    {\textsc{Electronics}}&H@10&M@10&N@10&H@10&M@10&N@10\\
    \midrule
    MMSC &\textbf{0.989}&\textbf{0.816}&\textbf{0.859}&\textbf{0.978}&\textbf{0.821}&\textbf{0.860}\\
    w/o SSL (\Cref{eqn:self_supervised}) &0.981&0.797&0.844&0.969&0.774&0.822\\
    w/o TA (\Cref{sec:task_level})&0.980&0.798&0.844&0.970&0.787&0.832\\
    w/o SSL \& TA &0.983&0.789&0.837&0.959&0.749&0.800\\
    w/o MM (\Cref{sec:multi_modal})&0.981&0.772&0.824&0.965&0.787&0.831\\
    w/o BM (\Cref{sec:behavior_based})&0.891&0.622&0.687&0.732&0.487&0.545\\
    w/o $\mathcal{E}_{\text{LLM}}$ (\Cref{sec:multi_task})&0.988*&0.812*&0.856*&0.970*&0.789*&0.833*\\
    w/o 3rd-hop (\Cref{sec:implementation}) &0.981&0.785&0.834&0.966&0.785&0.829\\
    w/o 3rd-hop, $\mathcal{E}_{\text{LLM}}$&0.976&0.780&0.828&0.972&0.788&0.830\\
    \bottomrule
    \end{tabular} 
    \caption{Ablation Results on Office and Electronics. w/o SSL means the model was trained without the self-supervised learning objective. TA corresponds to the task-level embedding aggregation in ~\Cref{sec:task_level}. MM denotes the multi-modal learning component in ~\Cref{sec:multi_modal}. BM denotes the behavior-based learning component in ~\Cref{sec:behavior_based}, and $\mathcal{E}_{\text{LLM}}$ denotes the LLM-assisted supervision in ~\Cref{sec:multi_task}. w/o 3rd-hop (\cref{sec:implementation}) means we constrain the length of meta-paths to be less than 3 (\ie only 2nd-hop neighborhood).}
    
    \label{tab:ablation}
    \end{table}
\subsection{Substitutable and Complementary Relationship Inference (RQ1)}
\label{subsec:sub_com_rec_results}

We report the results in~\Cref{tab:amazon-results}.
\name achieves a significant average improvement of 26.1\% in M@10 for substitutable and 39.2\% for complementary inference, respectively.
DHGAN~\cite{dhgan} is the strongest baseline for substitutability, while GAT~\cite{gat} is competitive for complementarity on several datasets.
Multi-modal foundation models (BLIP-2 and CLIP) perform noticeably worse, indicating that content-only encoders without relational learning are insufficient for item-item relation inference.

We observe more substantial performance gains from the baselines on complementary inference. This is because the complementary inference is more challenging, due to higher label noise and broader semantic diversity of complementary relations.
By combining behavior-derived structure with multi-modal metadata and explicitly denoising weak supervision, \name better captures such complex relations.



\subsection{Ablation Study (RQ2)}
\label{subsec:ablation_study}
We conduct ablation studies to assess the contribution of each component of \textsc{MMSC} (\Cref{tab:ablation}). Removing the self-supervised objective (w/o SSL) consistently degrades performance, especially for complementary inference, confirming its effectiveness in denoising noisy behavior-derived relations. Removing task-level aggregation (TA) also hurts performance, suggesting that controlled sharing between substitutable and complementary representations is beneficial.

Both the multi-modal module (MM) and behavior module (BM) contribute substantially to performance. The larger drop from removing BM indicates that behavior-based learning captures fine-grained relational structure, while MM provides complementary semantic information from item content. LLM-assisted supervision further improves performance, with larger gains on Office; the smaller gain on Electronics may be due to using the same 500K LLM-labeled pairs for a much larger catalog, which we further analyze in \Cref{sec:llm_augmentation_analysis}. Removing 3-hop meta-paths hurts performance, showing that higher-order connectivity provides useful evidence beyond direct neighbors. Finally, the joint ablation removing both 3-hop paths and $E_{\mathrm{LLM}}$ performs worse than removing only 3-hop paths, indicating that LLM filtering remains useful even without deeper meta-paths.
\renewcommand*{\factor}{0.085}
\begin{table}[h]
    \centering
    \small
    \begin{tabular}{p{0.15\linewidth}p{\factor\linewidth}p{\factor\linewidth}p{\factor\linewidth}p{\factor\linewidth}p{\factor\linewidth}
        p{\factor\linewidth}}
         & \multicolumn{3}{c}{\textsc{substitutable}}& \multicolumn{3}{c}{\textsc{Complementary}}\\
    \toprule
    {\textsc{Office}} &H@10&M@10&N@10&H@10&M@10&N@10\\
    \midrule
    GAT & 0.188 & 0.078 & 0.104 & 0.049 & 0.022 & 0.028 \\
    DHGAN &0.247 &0.100 &0.135 &0.143 &0.064 &0.082 \\
    BLIP-2-SA & 0.834* & 0.532* & 0.604* & 0.702* & 0.435* & 0.499* \\
    \midrule
    \name & \textbf{0.921} & \textbf{0.702} & \textbf{0.754} & \textbf{0.775} & \textbf{0.505} & \textbf{0.568} \\
    \% Improv & +10.4\% & +32.0\% & +24.8\% & +10.4\% & +16.1\% & +13.8\% \\
    \toprule
    {\textsc{Electronics}}      &H@10&M@10&N@10&H@10&M@10&N@10\\
    \midrule
    GAT & 0.161 & 0.066 & 0.088 & 0.080 & 0.025 & 0.038 \\
    DHGAN &0.224&0.106&0.133&0.183&0.093&0.113 \\
    BLIP-2-SA & 0.814* & 0.495* & 0.571* & 0.550* & 0.302* & 0.360* \\
    \midrule
    \name & \textbf{0.922} & \textbf{0.698} & \textbf{0.753} & \textbf{0.687} & \textbf{0.416} & \textbf{0.480} \\
    \% Improv & +13.3\% & +41.0\% & +31.9\% & +24.9\% & +37.7\% & +33.3\% \\
    \bottomrule
    \end{tabular}
    \caption{Cold-start Results on Office and Electronics.}
    \label{tab:amazon-office-cold}
        \vspace{-10pt}
    \end{table}
\subsection{Cold-start Inference (RQ3)}
\label{subsec:cold_start_sub_com_rec_results}
\subsubsection{\textbf{Cold-start Inference Procedure}}
\label{sec:cold_start}
We explore the effectiveness of \name under cold-start settings.
We evaluate cold-start items $\mathcal{V}'$ (\ie items that do not appear in training and have no observed behavior edges.), where  $\mathcal{V} \cap \mathcal{V}' = \emptyset$ and $\nexists\ \mathcal{E}_{ij'}, v_i \in \mathcal{V} \text{ and } v_{j'} \in \mathcal{V}'$. We sample 20\% of items as cold-start items and remove all their edges from the training set.
We obtain the initial item representation $\mH' = \mathcal{M}(X_i')$ for cold-start items using the multi-modal foundation models. Then, we use $\mH'$ as a query and search $\mH$ for the top-k most similar items (denoted as $\mathcal{C}$) in the existing item inventory, \ie items that appeared in the training set. Then, we mean pool the final representations of the selected items to obtain the final representation of items in $\mathcal{C}$. The final representation of the cold-start item is:
\begin{equation}
    \ve_j' = \frac{1}{|\mathcal{C}_{j'}|} \sum_{v_i \in \mathcal{C}_{j'}} \ve_i, \quad \forall v_{j'} \in \mathcal{V}'
\end{equation}
\label{eq:cold_start}
Note that all model parameters are fixed during cold-start inference. We adapt GAT and DHGAN using the same inference procedure.

\subsubsection{\textbf{Cold-start Results}}
We present the results in~\Cref{tab:amazon-office-cold}. Notably, \name significantly outperforms baselines by an average of 36.5\% (substitutable) and 26.9\% (complementary). GAT and DHGAN perform poorly in cold-start scenarios since their reliance on graph homophily limits generalizability to disconnected items. While BLIP-2 adapts better to cold-start scenarios through multi-modal information, it still underperforms compared to \textsc{MMSC}, suggesting multi-modal information alone might not be sufficient for cold-start scenarios and highlighting the advantage of \name's combined use of user behavior and content information in cold-start settings.
\begin{figure}[h]
    \centering
    \includegraphics[width=\linewidth]{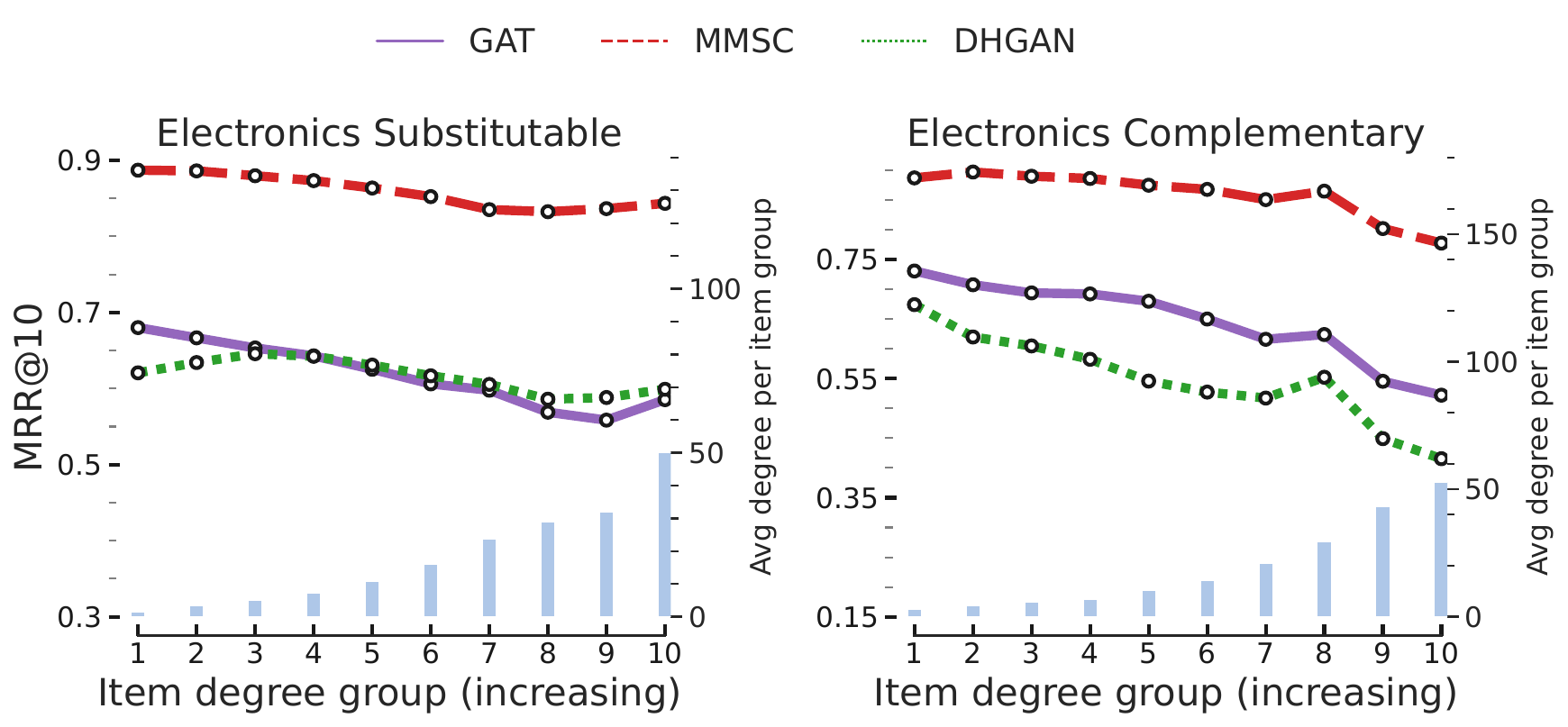}
    \caption{Performance on Electronics w.r.t. different item degree groups. \name shows greater improvement on items with less behavior data (Group 1-3) in substitutable inference and items with more behavior data (Group 8-10) in complementary inference.}
    \label{fig:qual_com}
\end{figure}
\begin{figure}[h]
    \centering
    \includegraphics[width=\linewidth]{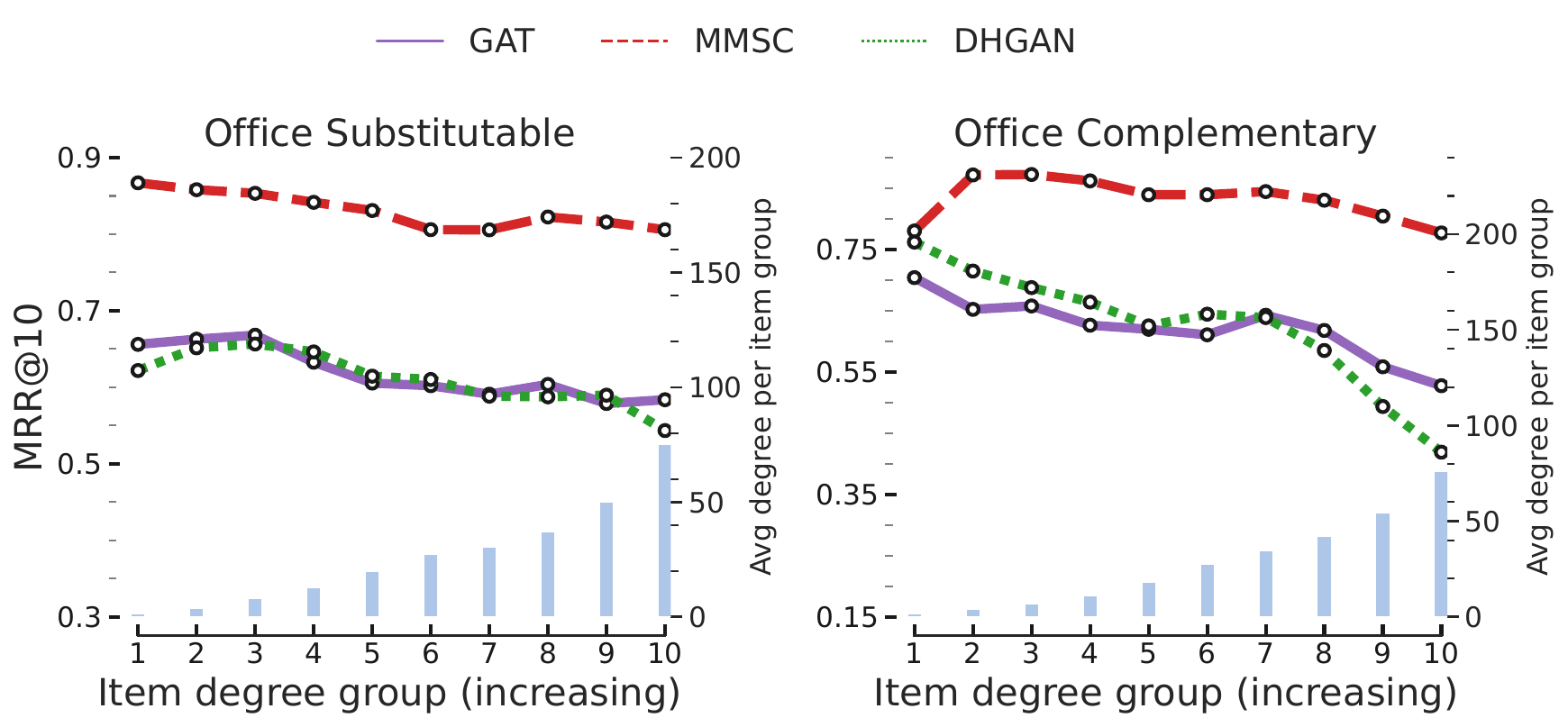}
    \caption{Performance on Office w.r.t. different item degree groups. Similarly, \name shows greater improvement on items with less behavior data (Group 1-3) in substitutable inference and items with more behavior data (Group 8-10) in complementary inference.}
    \label{fig:qual_sub}
\end{figure}
\subsection{Qualitative Analysis (RQ4)}
\label{subsec:qualitative_analysis}

\subsubsection{\textbf{Performance Across Item Degree Groups}}
\label{sec:degree_groups}
To understand how \name behaves under varying levels of behavioral supervision, we stratify test cases by the degree of the source item in the behavior-derived item-item graph. Specifically, we sort items by degree and partition them into 10 equal-sized groups (Group 1: lowest degree; Group 10: highest degree). The blue bars in \Cref{fig:qual_com,fig:qual_sub} show the average degree per group.

Overall, \name consistently outperforms all baselines across the degree spectrum for both substitutable and complementary inference. For substitutable inference, we observe the largest relative improvements in low-degree groups (Groups 1-3), where behavioral neighborhoods are sparse and graph-only models have limited relational evidence. In contrast, for complementary inference, baseline performance tends to deteriorate more noticeably in higher-degree groups (\eg Groups 8-10), likely due to the increased heterogeneity and noise among co-purchased edges. \name remains comparatively stable, leading to larger margins in these groups.

\begin{figure}[t]
    \centering
    \includegraphics[width=\linewidth]{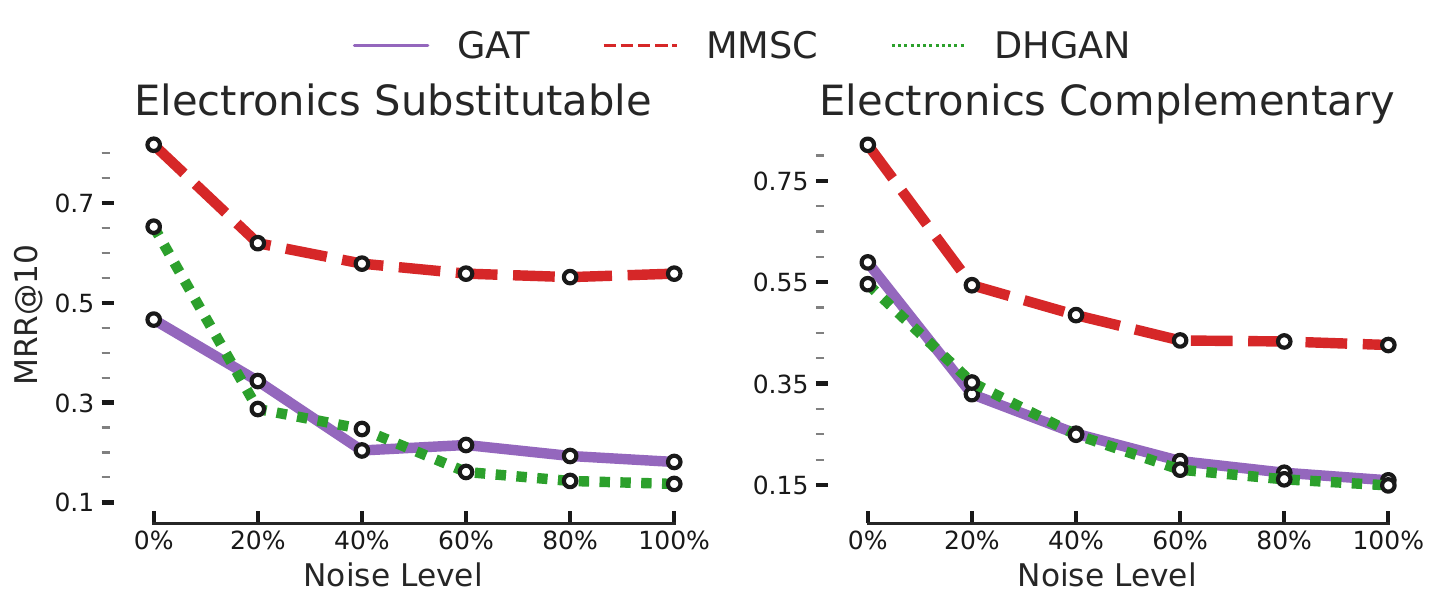}
    \caption{Performance on Electronics w.r.t. different noise levels.}
    \label{fig:noise_com}
\end{figure}
\begin{figure}[t]
    \centering
    \includegraphics[width=\linewidth]{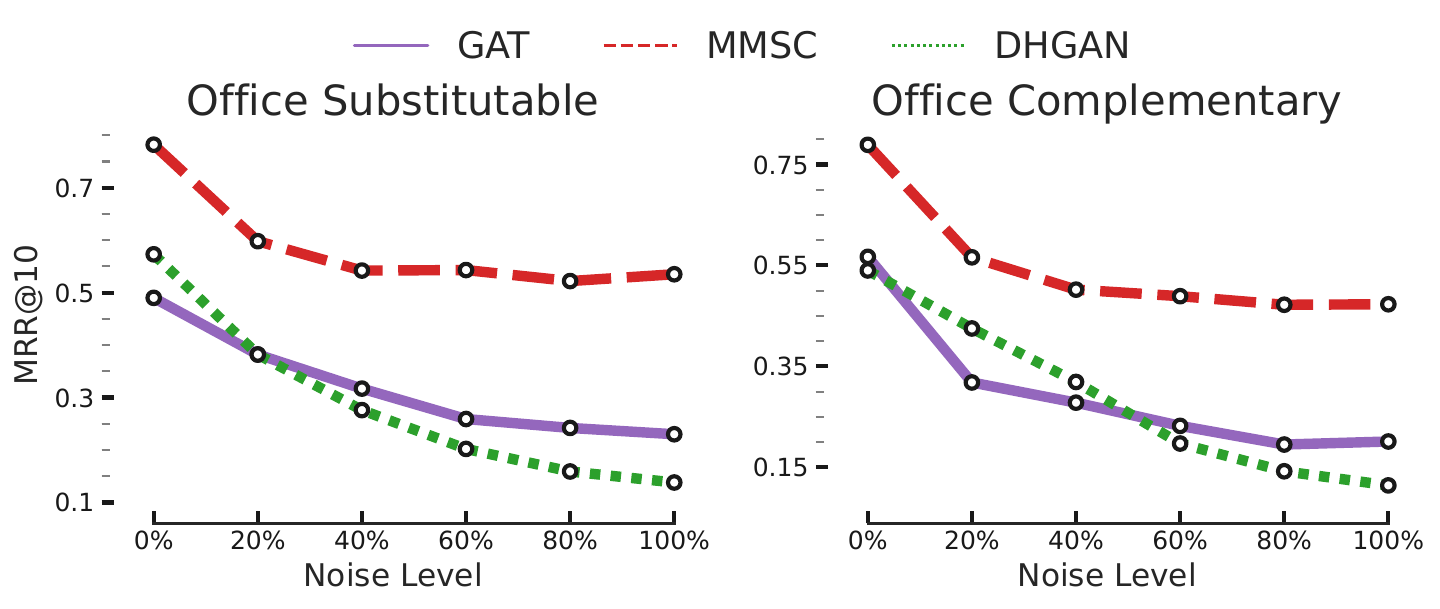}
    \caption{Performance on Office w.r.t. different noise levels.}
    \label{fig:noise_sub}
    \vspace{0pt}
\end{figure}
\subsubsection{\textbf{Robustness to Noise}}
We study how robust \name is to noise in user behaviors (\Cref{fig:noise_com} and \Cref{fig:noise_sub}). We evaluate robustness by injecting random (non-existing) behavioral edges into the item-item graph.
A noise level of $0\%$ uses the original graph, while $100\%$ injects the same number of random edges as the number of original edges.
As the noise level increases, \name maintains more stable performance and the relative improvement over baselines widens in both inference tasks, suggesting that \name is more resilient to spurious behavior-derived connections.

\begin{table}[h]
\centering
\small
\begin{tabular}{llccc}
\toprule
\textsc{Dataset} & \textsc{Source} & \textsc{Sub. Pos.} & \textsc{Comp. Pos.} & \textsc{Tail Cov.} \\
\midrule
\multirow{3}{*}{\textsc{Office}}
& Direct & 78.7\% & 24.4\% & 1.7\% \\
& 2-hop  & 26.1\% & 15.7\% & 4.3\% \\
& 3-hop  & 4.2\% & 8.3\% & 8.8\% \\
\midrule
\multirow{3}{*}{\textsc{Electronics}}
& Direct & 81.6\% & 35.2\% & 2.1\% \\
& 2-hop  & 22.3\% & 13.5\% & 7.4\% \\
& 3-hop  & 3.2\% & 6.0\% &8.9\% \\
\bottomrule
\end{tabular}
\caption{Meta-path depth and noise-coverage trade-off. For each source and relation type, we sample 1,000 candidate pairs and apply the same LLM verifier used in preprocessing. Sub. Pos. and Comp. Pos. denote the fraction of sampled pairs judged as valid substitutable and complementary relations, respectively. Tail Cov. denotes the fraction of tail items covered by the candidate pool.}
\label{tab:metapath_noise}
\end{table}


\subsubsection{\textbf{Meta-path Depth and Noise-Coverage Trade-off}}
To analyze whether deeper meta-paths improve coverage at the cost of introducing noisier candidates, we sample candidate pairs from direct edges, 2-hop paths, and 3-hop paths, and verify them using the same LLM filter as in preprocessing. As shown in \Cref{tab:metapath_noise}, direct edges have the highest positive rate but cover fewer tail items, while deeper meta-paths improve tail coverage but yield lower LLM-positive rates. This suggests a precision-coverage trade-off: deeper meta-paths recover additional sparse-item relational evidence, but require denoising. This also explains the ablation results in \Cref{tab:ablation}, where removing 3-hop paths hurts performance, while LLM-assisted supervision remains beneficial for filtering noisy behavior-derived candidates.

\begin{figure}[h]
    \centering
    \includegraphics[width=\linewidth]{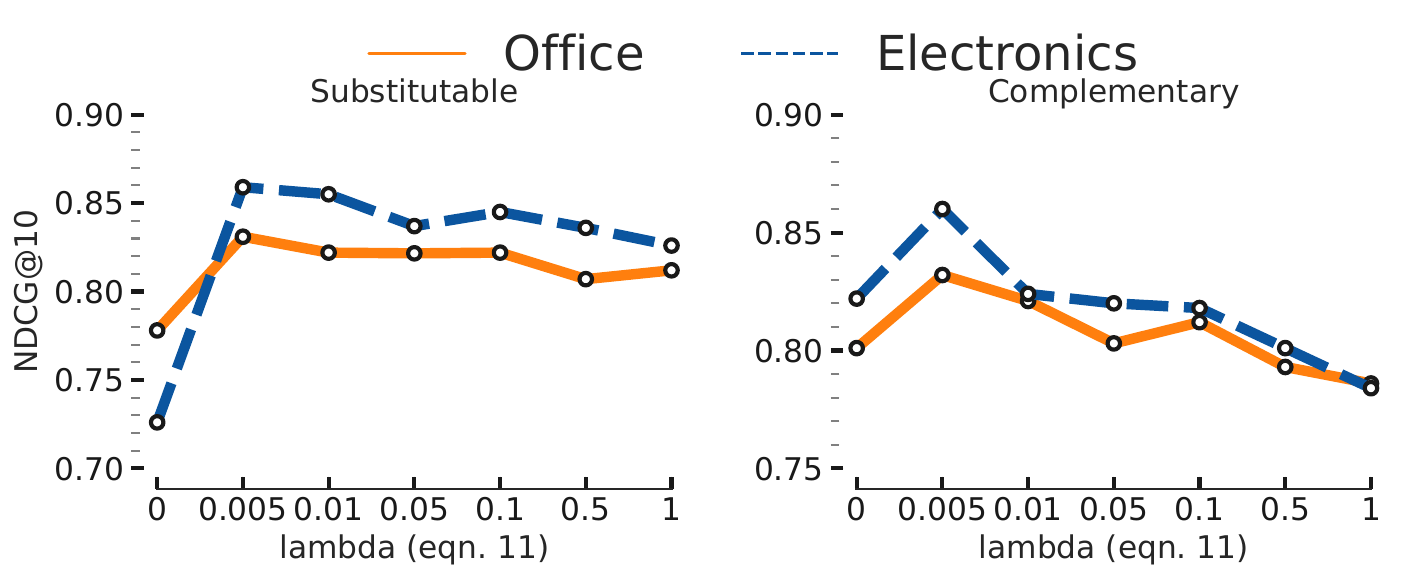}
    \caption{Performance of different $\lambda$ in ~\Cref{eq:final_loss}.}
    \label{fig:lambda-com}
\end{figure}

\vspace{-5pt}
\subsection{Sensitivity Analysis (RQ5)}
\label{subsec:parameter_sensitivity_analysis}

\subsubsection{\textbf{Sensitivity to $\lambda$ (\Cref{eq:final_loss})}}
\name achieves the best performance at $\lambda=0.005$ (\Cref{fig:lambda-com}), with accuracy improving as $\lambda$ increases up to this point.
This trend indicates that a moderate weight on the self-supervised term yields a favorable trade-off between denoising and supervised optimization, whereas overly large $\lambda$ can dominate training and hurt performance.
We also find that substitutable inference is less sensitive to $\lambda$ than complementary inference, implying that complementary inference benefits more from careful tuning.

\begin{figure}[h]
    \centering
    \includegraphics[width=\linewidth]{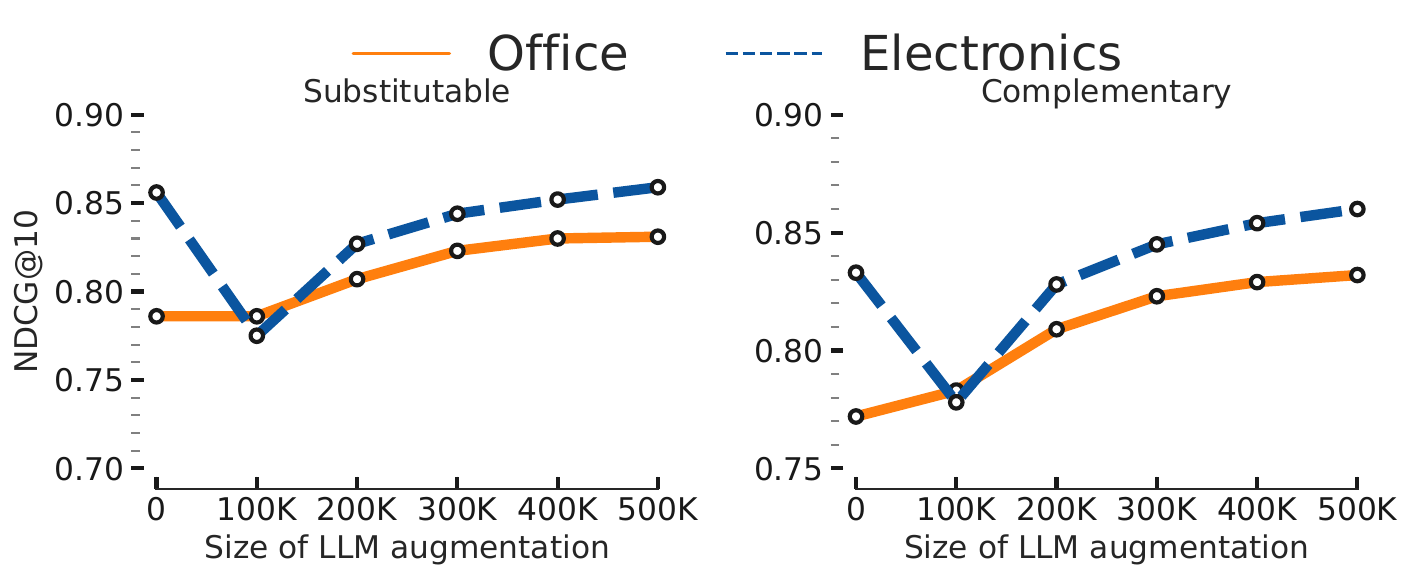}
        \caption{Performance varying size of $\mathcal{E}_{\text{LLM}}$ in \Cref{sec:llm_augmented}.}
    \label{fig:qual_llm}
\end{figure}
\subsubsection{\textbf{Sensitivity to $\mathcal{E}_{\text{LLM}}$ (\Cref{sec:llm_augmented})}} \label{sec:llm_augmentation_analysis}
We vary $|\mathcal{E}_{\text{LLM}}|$ from 100K to 500K (\Cref{fig:qual_llm}); $|\mathcal{E}_{\text{LLM}}|=0$ corresponds to training without LLM-assisted supervision (consistent with the ablation in~\Cref{tab:ablation}).
Performance generally improves as $|\mathcal{E}_{\text{LLM}}|$ increases.
On Office, LLM-assisted supervision yields consistent gains across sizes.
On Electronics, we observe non-monotonic behavior at smaller sizes (100K and 200K), which may indicate that limited augmented supervision is insufficient for a larger and more heterogeneous catalog.
Notably, LLM-assisted supervision can reduce the amount of training supervision while improving its quality, which is desirable in large-scale settings.
\vspace{-5pt}
\subsection{Discussion}
\label{subsec:discussion}
\name achieves strong improvements over competitive baselines for substitutable and complementary inference, and remains effective in cold-start scenarios.
It also consistently outperforms baselines across item degree groups and under increasing levels of injected noise, suggesting that \name learns more robust item representations from noisy, weak behavior supervision.
These results indicate that \name can be particularly useful in practical settings where behavioral signals are both noisy and long-tailed.
\paragraph{\textbf{Limitations:}}
First, \name assumes static relationships and does not model temporal dynamics (\eg seasonality or evolving complements).
Second, \name scores item pairs primarily through shared item representations and does not incorporate edge-specific context (\eg query intent, user segment, or session context), which could be important for capturing certain types of complementarity. Finally, \name does not consider independent item-item relationships during training and inference.
We leave these directions to future work.

\vspace{-8pt}
\section{Conclusion}
\label{sec:conclusion}
We identify two key challenges in modeling substitutable and complementary item relationships: (i) behavior-derived edges (\eg co-view/co-purchase) are noisy proxies for true relations, and (ii) user behaviors follow heavy-tailed distributions, leaving many items with sparse relational evidence. To address these challenges, we propose \textsc{MMSC}, a multi-modal relational item representation learning framework that jointly leverages behavior-based item associations and multi-modal item content, while denoising weak behavior supervision via self-supervised learning and LLM augmentation.

For future work, we will explore modeling independent item-item relations beyond co-occurrence graphs and incorporating temporal dynamics to capture how substitutable and complementary relationships evolve over time.

\vspace{-5pt}
\bibliographystyle{ACM-Reference-Format}
\bibliography{main,hs}

\clearpage

\end{document}